\documentstyle[prb,eqsecnum,aps,epsf]{revtex}

\begin{document}
%\draft

\twocolumn[\hsize\textwidth\columnwidth\hsize\csname@twocolumnfalse%
\endcsname

\title{Interacting electrons in magnetic fields:\\
       Tracking potentials and Jastrow-product wavefunctions}
\author{G\'{a}bor F\'{a}th\cite{address1} and Stephen B. Haley\cite{address2}}
\address{Institute of Theoretical Physics,
University of Lausanne,
Ch-1015 Lausanne, Switzerland}

\date{15 January 1998}
\maketitle

\begin{abstract}
The Schr\"{o}dinger equation for an interacting spinless electron
gas in a nonuniform magnetic field admits an exact solution in Jastrow
product form when the fluctuations in the magnetic field track the
fluctuations in the scalar potential. For tracking realizations in a
two-dimensional electron gas, the degeneracy of the lowest Landau level
persists, and the ``tracking'' solutions span the ground state subspace.
In the context of the fractional quantum Hall problem, the Laughlin wave
function is shown to be a tracking solution. Tracking solutions for
screened Coulomb interactions are also constructed. The resulting
wavefunctions are proposed as variational wave functions with potentially
lower energy in the case of non-negligible Landau level mixing than the
Laughlin function.
\end{abstract}

\vskip 0.3 truein
]

\section{Introduction}

Modern semiconductor technology has produced exceedingly pure 
materials, and nanostructures that have a number of interesting 
properties at low temperatures. A comprehensive review is given by 
Beenakker and van Houten.\cite{beenakker}   Modulation-doped 
GaAs-Al$_x$Ga$_{1-x}$As heterojunctions with precisely defined narrow 
potential wells at the interface strongly quantize electronic motion 
pependicular to the interface, creating a two dimensional electron gas 
(2-DEG).  If a magnetic field $B_a$ is applied perpendicular to the 
interface, electronic motion is further quantized in the interface 
plane, giving rise to a series of condensed Landau energy levels. At 
high magnetic field values the Hall conductance is accurately and 
robustly quantized in integer multiples of $e^2/h$. The resulting
``integer quantum Hall effect'' can be understood to a large extent within a
one-electron picture. On the other hand, quantization of the Hall
conductance at fractional filling factors, the ``fractional quantum Hall
effect'' (FQHE), is intrinsically a many-body effect in which
electron-electron interactions play the principal role.

Near Landau filling factors $\nu = n/(2mn +1)$, with $n,m$ integer,
the Hamiltonian of the strongly interacting 2-DEG 
can be transformed to a weakly interacting composite-fermion (CF) 
model in which each electron is coupled to an even number of magnetic
flux quanta $\tilde{\phi} = 2m$. Since its introduction, the CF
model\cite{jain} has attracted considerable theoretical
attention.\cite{lopez,halperin,brey,chklovskii,read,simon,kopietz}
The essence of the electron-electron correlations is captured by the
single-particle mean field approximation for the CF's, which are subject
to a reduced magnetic flux density $B_{\rm eff} = B_a(1-\tilde{\phi}\nu)$.
Since the CFs observe a reduced mean magnetic field, they fill
up a modified Landau level structure, in which the degeneracy of each level
is reduced appropriately. As a result, the weakly interacting CF gas is
described by an effective filling factor $\nu_{\rm eff}=\nu B_a/B_{\rm eff}$
which becomes integer $\nu_{\rm eff}=n$ at the above special fractional $\nu$ 
values.
The fractional quantum Hall effect follows by assuming that the CF's manifest
the integer quantum Hall effect when $\nu_{\rm eff}$ is integer.

The composite fermion (and similar composite boson) theories of the fractional
quantum Hall effect were largely motivated by the success of Laughlin's
variational wavefunction.\cite{laughlin1,laughlin2} Laughlin's wavefunction 
attaches collective vorticity to the electron gas which efficiently reduces the 
probability of two electrons approaching each other, and the Coulomb energy is 
reduced when $\nu$ is a simple odd fraction $\nu=1/(2m+1)$.
It was soon realised that such vorticity can be generated by a
singular gauge transformation which attaches infinitesimal flux-tubes to the 
particles, i.e., transforms electrons into composite objects with transmuted 
statistics.\cite{lopez}  However, such singular gauge transformations by 
themselves are not capable of reproducing the full Laughlin wavefunction, since 
they do not give its correct radial dependence. The full Laughlin wavefunction 
is obtained by taking into account fluctuation effects around the mean field 
solution.\cite{lopez}

The Laughlin wavefunction is a trial state completely formed within
the subspace of the {\em lowest} Landau level. It is believed to reflect the 
most
important aspects of the FQHE ground state at odd fractions, $\nu=1/(2m+1)$,
in the high magnetic field limit, where Landau level mixing can be neglected. 
Similarly, suggested generalizations to other fractional filling factors such as
$\nu=n/(2mn+1)$ are appropriately projected onto the lowest Landau 
level.\cite{jain}  In the ground state it is a priori not clear how to take a 
strong electron-electron interaction into consideration, or alternatively a 
weaker magnetic field, when the mixing of the Landau levels is not negligible. 
In such cases the 
true many-body ground state also has components from higher Landau levels. 

In this paper we develop exact eigenfunctions $\Psi$ of the Schr\"{o}dinger
equation for an interacting electron gas in non-uniform magnetic fields.
The initial insight and motivation is taken from the exactly solvable 1D
constructions of Calegero\cite{calogero} and Sutherland\cite{sutherland},
Polychronakos\cite{polychronakos}, and Haley\cite{haley} in the absence
of magnetic fields, and from the 2D constructions for the Pauli Hamiltonian
introduced by Aharonov and Casher\cite{aharonov}, and further developed by
Dubrovin and Novikov\cite{dubrovin} for an electron in a periodic magnetic 
field.  Our analysis is based on an ansatz that linearizes the vector Riccati
equation in ${\bf S} = -\nabla\ln\Psi$, which is equivalent to the 
Schr\"{o}dinger equation.  Our linearization is valid in any
number of dimensions, but it is most easily utilized in two dimensions.
We refer to the ansatz based solutions as ``tracking'' solutions, since they
satisfy certain linear first order differential equations containing the
vector potential, subject to a constraint relating the vector and scalar 
potentials.  We demonstrate that the Laughlin wave function 
$\Psi_L$, and any other wavefunction completely in the subspace of the lowest 
Landau level, is a tracking solution.  This tracking solution utilizes the 
Chern-Simons vector potential ${\bf a}_{cs}$, but in a different way than used 
in the composite fermion theories: In our theory, ${\bf a}_{cs}$ 
does not appear in the Hamiltonian.

We also derive a new trial wave function $\Psi_C$ that is the exact tracking
solution for the screened Coulomb potential.  A priori, $\Psi_C$ is just as 
close to the real physical situation as $\Psi_L$: While $\Psi_L$ misses the 
correct treatment of the Coulomb interaction, $\Psi_C$ requires the presence of 
a two-body magnetic field that tracks the scalar potential. The advantage of 
$\Psi_C$, however, is that when it is used as a trial wavefunction, it depends 
on free variational parameters and thus potentially gives a lower energy than 
$\Psi_L$, which has no adjustable parameters.  The wavefunction $\Psi_C$ is 
proposed to improve the FQHE ground state, especially in the case when Landau 
level mixing is important.

%=============================================================================
%\newpage
\section{GENERAL FORMULATION AND TRACKING}

Consider the time-dependent Schr\"{o}dinger equation for N electrons, each
with charge $-e < 0$ and effective mass $m^*$. An electron with coordinate
${\bf r}_i$ is subject to a scalar potential $V({\bf r}_i,t)$ and a vector
potential ${\bf A}({\bf r}_i,t)$, which in general include contributions from
both external fields and internal interactions depending on the coordinates
of all the electrons.  In this context, the Schr\"{o}dinger equation has the 
form 
\begin{eqnarray}
   \sum_{i=1}^N &&\left\{[\nabla_i + i{2\pi \over \phi_0}{\bf A}({\bf r}_i, t)]^2- 
   \gamma V({\bf r}_i, t)\right\}\Psi(\{{\bf r}\}, t) \nonumber\\
   &&= -i\gamma\hbar\frac{\partial}{\partial t}\Psi(\{{\bf r}\}, t) .
   \label{schrod}
\end{eqnarray}
where $\{{\bf r}\} = \{{\bf r}_1, {\bf r}_2, ...{\bf r}_i, ...{\bf r}_N\}$, 
and $\nabla_i $ is the gradient with respect to the electron
coordinate ${\bf r}_i$. The constant $\gamma = 
2m^*/\hbar^2$, and $\phi_0 = h/e$ is the flux quantum. Particles are labeled by 
Latin subscripts, but the constant $i = \sqrt{-1}$.  To assure the hermiticity 
of the Hamiltonian, the vector potential ${\bf A}({\bf r}_i,t)$ is
assumed to be {\em real}.  
The scalar potential has the form
\begin{equation}\label{coul}
   V({\bf r}_i,t) = V_a({\bf r}_i,t)+
                    \sum_{j\ne i} V_{\rm c}({\bf r}_i - {\bf r}_j),
\end{equation}
where $V_a({\bf r}_i,t)$ is a time-dependent applied scalar potential,
and $V_{\rm c}$ is an internal two-body electron-electron interaction
potential, such as the Coulomb interaction.

%-------------------------------------------------------------------------- 
\subsection{Riccati Equation}
The starting point of the analysis is a reformulation of the Schr\"{o}dinger 
equation (\ref{schrod}) in terms of a function $Q(\{{\bf r}\},t)$ defined by
\begin{equation}\label{Psi}
   \Psi(\{{\bf r}\}, t) = \exp[-Q(\{{\bf r}\}, t)].
\end{equation}
In general, $Q=-\ln \Psi$ is a complex multivalued function.
% which requires some mathematical rigor in the analysis.
Substituting $\Psi$ from Eq.\ (\ref{Psi}) into Eq.\ 
(\ref{schrod}) leads to the nonlinear 
vector Riccati equation
\begin{eqnarray}
   \sum_{i=1}^N\Big[&&\nabla_i\cdot ({\bf S}_i - i{2\pi \over \phi_0}{\bf A}_i)
   - ({\bf S}_i - i{2\pi \over \phi_0}{\bf A}_i)\cdot
   ({\bf S}_i - i{2\pi \over \phi_0}{\bf A}_i) \nonumber\\
   &&+ \gamma V_i\Big] \Psi(\{{\bf r}\},t)=  
   - i\gamma\hbar\frac{\partial}{\partial t}Q(\{{\bf r}\}, t)\,
   \Psi(\{{\bf r}\},t)
   \label{schrodQ}
\end{eqnarray}
where
\begin{equation}
   {\bf S}_i = {\bf S}_i(\{{\bf r}\}, t) = \nabla_i Q(\{{\bf r}\}, t),
   \label{SQ}
\end{equation}
${\bf A}_i = {\bf A}({\bf r}_i, t)$, and $V_i = V({\bf r}_i, t)$.  The above form of the 
Riccati 
equation contains $\Psi$ as a multiplicative factor on both sides. In the 
following this factor will be divided out, even though such division is not 
legitimate at the zeros of $\Psi$.   This subtlety results in the appearance of 
spurious delta-function potentials at the position of the zeros.  
However, as shown in section II, the correct physics is preserved.

Although the original Schr\"odinger equation is a linear, second order
equation for $\Psi$, Eq.\ (\ref{schrodQ}) is a first order, nonlinear
equation for the derivatives of $Q$.
Given real vectors ${\bf A}_i$ and real scalars $V_i$,
Eq.\ (\ref{schrodQ}) can, in principle, be solved for $Q(\{{\bf r}\},t)$. 
For time-independent potentials, the 
function $Q(\{{\bf r}\},t) = Q(\{{\bf r}\}) + i(E/\hbar)t$, and 
Eq.\ (\ref{schrodQ}) becomes the time-independent energy 
eigenvalue equation
\begin{eqnarray}
   \sum_{i=1}^N &&\left[\nabla_i\cdot ({\bf S}_i 
                     -i{2\pi \over \phi_0}{\bf A}_i)\right. \nonumber\\
   &&\left. - ({\bf S}_i - i{2\pi \over \phi_0}{\bf A}_i)\cdot
   ({\bf S}_i - i{2\pi \over \phi_0}{\bf A}_i) + \gamma V_i\right] = 
   \gamma E.  \label{eigS}
\end{eqnarray}
In the following we restrict our attention to time-independent problems, but it
will be noted that the tracking construction introduced in subsection II.B also 
linearizes Eq.\ (\ref{schrodQ}).

In one dimension the transformation defined by Eq.\ (\ref{Psi}) and 
(\ref{SQ}) leads to a scalar Riccati equation that is quite useful for 
obtaining the stationary, one-particle states for many scalar 
potentials in the absence of 
magnetic fields.\cite{haley}  In the many-particle case the major difficulty in 
solving Eq.\ (\ref{eigS}) is that the quadratic term in ${\bf S}_i$ may contain
interactions of order higher than those appearing in the scalar potential $V_i$.
To see this, consider wave functions in a Jastrow product form,
\begin{equation}\label{jastrow}
   \Psi(\{{\bf r}\}) = \prod_{i}\prod_{j>i} \Omega ({\bf r}_i-{\bf r}_j)\cdot
                     \prod_{i} \chi({\bf r}_i),                     
\end{equation}
where $\Omega({\bf r}_i-{\bf r}_j)$ is a two-body function depending
on the coordinate difference between the particles $i$ and $j$, and
$\chi_i=\chi({\bf r}_i)$ is a one-body function involving only the coordinates
of particle $i$. Due to the two-body factors $\Omega_{ij}$,
the vector ${\bf S}_i$ will be a sum over two-body functions, each of which
depends on the coordinates of the $i$th electron and another one; thus the 
quadratic term in ${\bf S}_i$ implicitly contains three body interactions.

One may conclude that a wavefunction in the Jastrow product form of 
Eq.\ (\ref{jastrow}) can only be an eigenstate if the model in question
contains an unphysical three-body potential.  Boson Hamiltonians of this kind 
have been investigated in Ref.\ \onlinecite{Kane}. 
There it is argued that even though the three-body term is explicitly present in
the Hamiltonian, it becomes irrelevant in the renormalization 
group sense; thus the Jastrow product, although it is not an 
exact eigenstate, describes correctly
the low-energy, long-wavelength physics of the remaining two-body Hamiltonian. 
While three-body terms are required in the 
generic case, Jastrow products can be exact eigenstates
of Hamiltonians {\em without} three-body terms, provided that special 
conditions are fulfilled, as shown in section II.B.

Well-known examples in one dimension are the Calogero-Sutherland 
models.\cite{calogero,sutherland,olshanetsky}
They form a class of integrable many-body systems,  
which describe particles interacting through two-body potentials $V_{\rm c}(
r_i - r_j)$ in zero magnetic field ${\bf A}_i\equiv 0$. Their ground state
can be written in the product form of Eq.\ (\ref{jastrow}). For such systems 
the sum of the quadratic terms of the associated 1D Riccati equation reduces
to a sum of two body interactions
\begin{equation}\label{suther}
   \sum_i S_i^2 =\sum_i\sum_{j>i} F_{ij}.  
\end{equation}
The function $F_{ij}=F(r_i-r_j)$ is symmetric, and only
depends on the coordinate difference between the $i$th and $j$th particles.
This equation is a consequence of a more elementary functional equation
(Sutherland's equation\cite{sutherland}) assuring that the scattering of three 
particles can be factorized into two-particle scatterings.  
The class of models satisfying Eq.\ (\ref{suther}) is however limited:  
Typically, the two-body potential $V_{\rm c}$ is constrained to 
diminish over short distances as $1/r_{ij}^2$.

The Riccati equation, Eq.\ (\ref{eigS}), provides an alternative description of the 
quantum mechanical system.  However, care should be taken since not every 
solution of the Riccati equation yields a valid, single valued quantum mechanical 
wavefunction (we do not consider anyons here). 
Assume that a solution ${\bf S}_i$ of Eq.\ (\ref{eigS}) is found. The
function $Q$ can then be obtained by inverting Eq.\ (\ref{SQ}), which is
tantamount to the line integral
\begin{equation}
   Q(\{{\bf r}\})=Q(\{{\bf r}_0\})+
   \int_{\displaystyle \{{\bf r}_0\}}^{\bf\displaystyle \{r\}}
   \{d{\bf r}^\prime\}\cdot \{{\bf S}(\{{\bf r}^\prime\})\}, 
   \label{lineint}
\end{equation}
where $\{{\bf S}\}=\{{\bf S}_1,{\bf S}_2,\dots,{\bf S}_N\}$, and 
$\{d{\bf r}\}=\{d{\bf r}_1,d{\bf r}_2,\dots,d{\bf r}_N\}$ are $DN$-component
supervectors ($D$ is the dimension of the space), and $Q(\{{\bf r}_0\})$ is a 
constant.  In the many-particle case the line integral is path dependent 
whenever the $DN$ dimensional vector field $\{{\bf S}\}$ is non-conservative, 
i.e., $\partial S_i^\alpha/\partial r_j^\beta-\partial S_j^\beta/\partial 
r_i^\alpha \ne 0$ for some $i,j=1,\dots,N$; $\alpha,\beta=1,\dots,D$.  
Since the wavefunction $\Psi$ must be single valued, the 
function $Q$ is determined only up to an additive term $i2\pi m$, with $m$ 
integer. Hence, we allow a set of well-defined, isolated, singularities of 
$\delta$-function type in any two-dimensional slice $(i\alpha,j\beta)$ of the 
$DN$ dimensional space.  Restricting our attention only to solutions in the 
Jastrow-product form, it is enough to require a constraint on 
$\nabla_i\times{\bf S}_i$, i.e., for slices in the same particle subspace 
with $i=j$, since then the 
special form in Eq.\ (\ref{jastrow}) assures that, at most, similar
$\delta$-function type singularities appear in other slices with $i\ne j$.

In 3D we allow discrete loops described by vectors ${\bf c}_k(s)$,
parametrized by the length $s$, on which $\nabla_i\times {\bf S}_i$ is
proportional to a $\delta$-function.  We write
\begin{equation}\label{singS}
   \nabla_i\times{\bf S}_i = -i2\pi \,\sum_k m_k
                            \delta\big({\bf r}_i-{\bf c}_k(s)\big)\;
                            \partial_s{\bf c}_k(s),
\end{equation}
where $m_k$ is an integer, and $\partial_s{\bf c}_k(s)$ is a unit vector
tangential to loop $k$ at the point $s$.  Although the singularity loops are 
independent of ${\bf r}_i$, they may depend on the coordinates of the other 
particles. By virtue of (\ref{singS}), the value of the line integral in Eq.\ 
(\ref{lineint}) is determined only up to the term $i 2\pi  \sum_k m_k n_k$, 
where the integer $n_k$ is the Gauss linking number (winding number in 2D) of 
the path linked by the $k$th singularity loop. Since $\sum_k m_kn_k$ is 
integer, Eq.\ (\ref{singS}) assures that the wavefunction $\Psi$ in 
Eq.\ (\ref{Psi}) is single valued.  In a 2D plane defined by the perpendicular 
unit vector $\hat{\bf z}$, the vectors ${\bf c}_k(s) \rightarrow {\bf c}_k$ 
are discrete points, and $\partial_s{\bf c}_k(s) \rightarrow \hat{\bf z}$. 

%-------------------------------------------------------------------
\subsection{Tracking Constructions}
In this subsection we construct a class of exactly solvable many-body
Hamiltonians in higher dimensions, which contain both scalar and vector
potentials. The many-body eigenfunctions can be written in the product form 
(\ref{jastrow}), with each term related to terms of an appropriate
decomposition of the vector potential.
Our construction is motivated by the need to linearize the Schr\"odinger
equation to eliminate the three-body interaction terms generated by
the quadratic terms in ${\bf S}_i$ and ${\bf A}_i$ in (\ref{eigS}).
Linearization may be achieved by requiring that
\begin{equation}\label{zerosquare}
  \sum_i \left( {\bf S}_i- i{2\pi \over \phi_0}{\bf A}_i\right)^2 = 0,
\end{equation}
which is a sufficient condition for the factorizability of the wavefunction. 
Equation (\ref{zerosquare}) is a stronger restriction than that in Eq.\ 
(\ref{suther}), since it requires $F_{ij} = 0$. In 1D with ${\bf A}_i\equiv 0$, 
realizations of Eq.\ (\ref{zerosquare}) are extremely limited since
this equation implies that 
the wavefunction is a coordinate independent constant. It is the presence of the
vector potential, and especially the higher dimensionality which make the 
constraint (\ref{zerosquare}) plausible. We show in the following that Eq.\ 
(\ref{zerosquare}) provides nontrivial, useful solutions to higher dimensional 
models containing vector potentials.

To manifest the implications of Eq.\ (\ref{zerosquare}) we decompose the complex 
vector ${\bf S}_i$ as ${\bf S}_i = \Re{\bf S}_i + i\Im{\bf S}_i$ and write out 
the 
real and imaginary parts of Eq.\ (\ref{zerosquare}), respectively as 
\begin{eqnarray}
  \sum_i\left[ (\Re{\bf S}_i)^2 - (\Im{\bf S}_i - {2\pi \over \phi_0}{\bf 
  A}_i)^2\right] &&= 0,\nonumber\\
  \sum_i \Re{\bf S}_i\cdot(\Im{\bf S}_i - {2\pi \over \phi_0}{\bf A}_i) = 0.&&
  \label{zerosquare1}
\end{eqnarray}
For a system of identical particles, each term in Eq.\ (\ref{zerosquare1}) is 
zero, and consequently the two real vectors $\Re{\bf S}_i$ and $\Im{\bf S}_i -
{2\pi /\phi_0}{\bf A}_i$
are orthogonal to each other and have the same length at any point of
the configuration space $\{{\bf r}\}$. Evidently, if only
${\bf A}_i$ is given, this restriction still allows an infinite number of
possible solutions for the complex vector ${\bf S}_i$. To parametrize these 
solutions we formally decompose ${\bf A}_i$ into the sum of two components
${\bf A}_i={\bf A}^{\rm I}_i+{\bf A}^{\rm II}_i$, with each term real.
Equation (\ref{zerosquare}), or 
equivalently the equations in (\ref{zerosquare1}), are satisfied by the ansatz
\begin{equation}\label{tr}
   {\bf S}_i = {2\pi\over \phi_0}\left[-{\hat{\bf n}\times
               {\bf A}^{\rm I}_i\over D_i(\hat{\bf n})} +
               i{\bf A}^{\rm II}_i \right],\qquad 
   D_i(\hat{\bf n}) = |\hat{\bf n}\times\hat{\bf A}^{\rm I}_i|,
\end{equation}
where $\hat{\bf n}$ is a unit vector with arbitrary direction, and the unit 
vector 
$\hat{\bf A}^{\rm I}_i = {\bf A}^{\rm I}_i/ A^{\rm I}_i$. Equation (\ref{tr}) 
completely specifies the vector ${\bf S}_i$ in terms of $\hat{\bf  n}$ and the 
decomposition of ${\bf A}$.  Using the ansatz (\ref{tr}), the Riccati Eq.\
(\ref{eigS}) reduces to 
\begin{equation}\label{trackc}
   \sum_{i=1}^N\left[-\mu_B \nabla_i\cdot\left({\hat{\bf n}\times
   {\bf A}^{\rm I}_i\over D_i(\hat{\bf n})} +
   i{\bf A}^{\rm I}_i  \right) + V_i\right] = E,  
\end{equation}
where $\mu_B=e\hbar/2m^*$ is the Bohr magneton.  In general, the  first term
in Eq.\ (\ref{trackc}) is a complicated nonlinear function of the angle
between $\hat{\bf n}$ and ${\bf A}^{\rm I}$. Considerable simplification
is achieved when $\hat{\bf n}$ and the coordinate system can be chosen such
that $D_i(\hat{\bf n})$ and $\hat{\bf n}$ are constants.  In 3D these conditions
are restrictive, but in 2D they can always be satisfied (See Section II).
Assuming that $D_i(\hat{\bf n})$ and $\hat{\bf n}$ are constant, Eq.\
(\ref{trackc}) reduces to the linear equations
\begin{equation}\label{track}
   \sum_{i=1}^N\left[{\mu_B\over D_i(\hat{\bf n})}\hat{\bf n}\cdot
   (\nabla_i\times {\bf A}^{\rm I}_i) + V_i\right] =  E,  
\end{equation}
and
\begin{equation}\label{coulgauge}
   \sum_{i=1}^N\nabla_i\cdot{\bf A}^{\rm I}_i  = 0.  
\end{equation}
Since the rhs of Eq.(\ref{track}) is a 
coordinate independent constant $E$, Eq.\ (\ref{track}) can only hold if the 
spatial variations of the component of the magnetic field ${\bf B}^{\rm I}({\bf 
r}_i) = \nabla_i\times {\bf A}^{\rm I}_i$ in the direction $\hat{\bf n}$ track 
those of the scalar potential.  Although Eqs.\ (\ref{track}) and 
(\ref{coulgauge}) give a restriction only on the sum of $N$ terms, in case of 
identical particles they hold identically for each particle $i$, and each vector 
potential ${\bf A}^{\rm I}_i$ is in the Coulomb gauge, $\nabla_i\cdot{\bf 
A}^{\rm I}_i  = 0$. 

By virtue of the single valuedness condition Eq.\ (\ref{singS}), whose
rhs is purely imaginary, it follows using Eq.\ (\ref{tr}) that ${\bf B}^{\rm 
II}_i = \nabla_i\times {\bf A}^{\rm II}_i$ is necessarily restricted to the form
\begin{equation}\label{fluxtube}
  {\bf B}^{\rm II}_i  = - \phi_0 \sum_k m_k \delta({\bf r}_i-{\bf 
  c}_k)\partial_s{\bf c}_k(s),  
\end{equation}
with $m_k$ an integer. Thus the construction (\ref{tr}) excludes 
regular magnetic fields in ${\bf  B}^{\rm II}_i$,
only allowing singular flux tubes proportional to $\delta$-functions.
It is evident that if we already possess a 
solution of the Schr\"odinger equation, we can construct a new Hamiltonian, with 
the associated new eigenfunction, by adding a term $\delta{\bf A}^{\rm II}_i$
to the vector potential and the associated term $\delta\Im{\bf S}_i$ to the
logarithmic derivative of the 
wavefunction, providing that they satisfy Eq.\ (\ref{tr}). The 
reader recognizes that the imaginary part of Eq.\ (\ref{tr}) is related to gauge 
transformations generated by real functions $\Lambda({\bf r}_i)$, which 
modify the vector potential as $\delta{\bf A}_i=\nabla\Lambda_i$ and the phase
of the wavefunction as $i(e/\hbar)\Lambda_i$. Note that Eqs.\
(\ref{tr}) and (\ref{fluxtube}) also allow {\em 
singular} gauge transformations, generated by {\em multivalued} functions 
$\Lambda_i$. Singular gauge transformations may lead to more substantial
modifications in the model than regular transformations by adding singular flux 
tubes to the system and thus eventually transmuting particle statistics. 
 
The single valuedness condition Eq.\ (\ref{singS}) also implies a restriction
on the type-I component ${\bf A}^{\rm I}_i$.  Using Eq.\ (\ref{tr}) in 
$\nabla_i\times \Re{\bf S}_i = 0$, with $D_i(\hat{\bf n})$ and $\hat{\bf n}$ 
constant, and 
$\nabla_i\cdot{\bf A}^{\rm I}_i = 0$, gives the constraint
\begin{equation}\label{singSconII}
  (\hat{\bf n}\cdot \nabla_i) {\bf A}^{\rm I}_i = 0.
\end{equation}
Thus ${\bf A}^{\rm I}_i$ must be constant in the $\hat{\bf n}$ direction.  
However, since $\nabla_i\times {\bf A}^{\rm I}_i$ is not restricted, regular 
magnetic fields are allowed in ${\bf B}^{\rm I}_i$.

Equation (\ref{track}) is the main result of our paper. It completely determines 
the decomposition of the magnetic flux density, and if a physical problem, 
defined 
by giving the functions $V_i$ or ${\bf B}^{\rm I}_i$, and $\hat{\bf n}$ admits 
tracking solutions then there is only one tracking solution up to possible gauge 
transformations.  Equation (\ref{track}) together with the constraints in Eqs.\ 
(\ref{coulgauge}) - (\ref{singSconII}) further imply that the $N$-particle 
Schr\"odinger equation in Eq.\ (\ref{schrod}) has an exact tracking solution 
with energy $E$ and eigenfunction
\begin{eqnarray}\label{solution}
   \Psi(\{{\bf r}\}) &=& e^{-\Re Q(\{{\bf r}\})} e^{-i\Im Q(\{{\bf r}\})}.
\end{eqnarray}
For identical particles, the problem reduces to a
one-particle problem in external fields and in fields created by
other particles. The $N$-particle wavefunction can be written as a
product of one-particle wavefunctions (or in some cases as a linear 
superposition of these), where the one-particle functions also contain
the coordinates of the other particles as parameters.

The tracking wave function $\Psi$ may be obtained by solving Eq.\ (\ref{tr}), which 
expresses the real and imaginary parts of ${\bf S}_i = \nabla_i Q = -\nabla_i \ln \Psi$ 
in terms of the ${\bf A}_i^{\rm I}$ and ${\bf A}_i^{\rm II}$ components, respectively, of 
the vector potential decomposition. However, $|\Psi|$ can  be expressed directly as a 
function of the scalar potential $V_i$: Assuming that the particles are identical
and noting that ${\bf A}$ is real, Eqs.\ (\ref{eigS}) and
(\ref{zerosquare}) yield Poisson's equation 
\begin{equation}\label{poisson}
   \nabla_i^2 \,\Re Q = \gamma (E/N-V_i),
\end{equation}
with the energy eigenvalue $E$ determined by the tracking constraint (\ref{track}).  
Since the tracking solution is a Jastrow product, we find that $\ln|\Psi| = - \Re Q$ is 
determined by the sum of quadratures over the scalar potential,
\begin{equation}\label{QGF}
   \Re Q(\{{\bf r}\}) = \gamma \sum_{i=1}^N \int d^D{\bf r}_i^\prime\;
           G_D({\bf r}_i,{\bf r}_i^\prime)\,
           \left[ E/N - V({\bf r}_i^\prime)\right],
   \label{quadrature}
\end{equation}
where $G_D({\bf r}_i,{\bf r}_i^\prime)$ is the $D$-dimensional Green's function.  
Although we do not use Eq.\ (\ref{QGF}) to obtain the tracking solutions in Section III, 
it is applicable and it may be useful in other problems.

It is of interest to calculate the current density corresponding to the tracking 
solution.  The electron current density associated with some solution $\Psi$ of 
the Schr\"{o}dinger equation is defined as
\begin{equation}\label{jdef}
   {\bf j}_i = -{e\hbar\over m^*}\Re\left[\Psi^*(-i\nabla_i + {e\over\hbar}{\bf 
   A}_i)\Psi\right] = {e\hbar\over m^*}|\Psi|^2\Im [{\bf v}_i].    
\end{equation}
where Eq.\ (\ref{Psi}) was used to obtain the latter form, and  ${\bf v}_i = 
{\bf S}_i - i{2\pi \over \phi_0}{\bf A}_i$ is a normalized complex velocity.  
Applying  Eq.\ (\ref{tr}), the velocity is ${\bf v}_i  = \Re{\bf S}_i - 
i(2\pi/\phi_0){\bf A}^{\rm I}_i$, yielding the current density
\begin{equation}\label{tv}
   {\bf j}_i = -{e^2\over m^*}|\Psi|^2 {\bf A}^{\rm I}_i.
\end{equation}
Since the current is gauge invariant, the singular type-II component ${\bf 
A}^{\rm II}_i$ does not contribute.

%=============================================================================
\section{Tracking in two dimensions}

In this section we restrict the interacting electron gas to two dimensions, and 
develop several examples of tracking solutions.  In 2D, with both ${\bf S}_i$ 
and 
${\bf A}_i$ restricted to the $x$--$y$ plane, we seek tracking solutions with 
$\hat{\bf n} = \hat{\bf z}$. Noting that $D_i(\hat{\bf z}) = 1$, in 2D the 
vector ${\bf S}_i$ is defined by the equation
\begin{equation}\label{tr2D}
   {\bf S}_i = \nabla_i Q = {2\pi\over \phi_0}\left[-\hat{\bf z}\times
   {\bf A}^{\rm I}_i + i{\bf A}^{\rm II}_i \right].
\end{equation}
The tracking constraint (\ref{track}) becomes 
\begin{equation}\label{track2D}
   \sum_{i=1}^N\left[\mu_B B^{\rm I}({\bf r}_i) + V({\bf r}_i)\right] =  E,  
\end{equation}
with $B^{\rm I}_i = \hat{\bf z}\cdot(\nabla_i\times {\bf A}^{\rm I}_i)$. The  
constraints (\ref{coulgauge}) - (\ref{singSconII}) are $\nabla_i\cdot {\bf 
A}^{\rm I}_i = 0$,
\begin{equation}\label{fluxtube2d}
   B^{\rm II}_i  = \hat{\bf z}\cdot (\nabla_i\times {\bf A}^{\rm II}_i) = - 
   \phi_0 \sum_k m_k \delta({\bf r}_i-{\bf c}_k),
\end{equation}
and $\partial {\bf A}^{\rm I}_i/\partial z = 0$, respectively.

In the following we assume that the particles are spinless fermions, and that 
the system is subject to an applied uniform magnetic field ${\bf B}_a=\hat{\bf 
z} B_a$, with $B_a>0$. The corresponding vector potential in the symmetric 
gauge, written in polar coordinates ${\bf r}=(r,\varphi)$, is
\begin{equation}\label{A_a}
   {\bf A}_a = {1\over 2}B_a r \hat{\varphi}.
\end{equation} 
We now develop a few interesting solutions of the tracking relation Eq.\  
(\ref{track2D}), relevant to the fractional quantum Hall effect, which
satisfy the constraints following Eq.\ (\ref{track2D}), and construct
the associated eigenfunctions.
          
%---------------------------------------------------
\subsection{Non-Interacting Electrons}

When ${\bf A}={\bf A}_a$ is the only potential which appears in the
Hamiltonian, i.e., the scalar potential $V=0$, the one-electron states form 
highly degenerate Landau levels. Restricting our attention to filling factors
$\nu=N\phi_0/B_a \le 1$, we consider exclusively the lowest Landau level (LLL),
which has the one-particle energy $E/N=\mu_B B_a$.  The single particle ($N=1$) 
Schr\"odinger equation with vector potential ${\bf A} = {\bf A}_a$ has   
eigenfunctions with anti-analytic prefactors of the form
\begin{equation}
   \psi_m({\bf r})\! =\! c_m \bar{Z}^m \exp\!\left(-{1\over 4} 
   l_a^{-2}Z\bar{Z}\right),    \quad Z\!=\!x\!+\!iy\!=\!re^{i\varphi},
   \label{basis}
\end{equation}
where $l_a^{-2} = eB_a/\hbar = (2\pi/\phi_0)B_a$, and $c_m \propto 
l_a^{-(m+3/2)}$ is a normalization constant. These one-particle basis functions 
$\psi_m$ with definite canonical angular momentum $m\hbar$ ($m$ a nonnegative 
integer) span the LLL. Substitution of $\psi_m$ into Eq.\ (\ref{schrod}) yields 
the corresponding energy eigenvalue $E_m = \mu_B B_a$. Using $\psi_m$ and ${\bf 
A}_a$ in Eq.\ (\ref{jdef}) gives the circulating current density 
\begin{equation}\label{j_m}
   {\bf j}_m = \mu_B |\psi_m|^2
               \left({2m\over r} - l_a^{-2}r \right)\hat{\varphi}
\end{equation}
Since $c_m^2 \propto B_a^{m+3/2}$, the current $j_m 
\rightarrow 0$ in the limit $B_a \rightarrow 0$, as expected when there is no
force on the electron.  

We now show that any basis state $\psi_m$ is a tracking solution corresponding 
to a 
$\delta$ function potential $V({\bf r},m)$.  Using the formalism of 
section II with $N=1$, we decompose the vector potential ${\bf A}_a$ as
${\bf A}_a = {\bf A}^{\rm I}+{\bf A}^{\rm II}$ with
\begin{equation}\label{AIAIIfp}
   {\bf A}^{\rm I} = {\bf A}_a - {\bf a}({\bf r}, m),\qquad
   {\bf A}^{\rm II} =  {\bf a}({\bf r}, m),
\end{equation}
where the particle index has been omitted, and ${\bf a}$ is a singular vector 
potential defined by
\begin{equation}\label{arm}
   {\bf a}({\bf r}, m)= {\phi_0\over 2\pi}{m\over r}\hat{\varphi}.
\end{equation}
The auxiliary vector potential ${\bf a}({\bf r}, m)$ attaches a fictitious flux 
tube 
of density
\begin{equation}
    b({\bf r}, m)= \hat{\bf z}\cdot\nabla\times {\bf a} =
                   m\phi_0 \delta({\bf r}),
\end{equation}
to the origin of the coordinate system, but since the flux tubes are attached 
with opposite sign in ${\bf A}^{\rm I}$ and ${\bf A}^{\rm II}$, they cancel out 
in the Hamiltonian and have no direct physical effect.

To form the corresponding tracking eigenfunction we solve Eq.\ (\ref{tr2D}),
with ${\bf A}^{\rm I}$ and ${\bf A}^{\rm II}$ given by Eq.\ (\ref{AIAIIfp}).
The imaginary part gives $\Im Q = m\varphi$, while the real part yields 
$\Re Q= 1/4 l_a^{-2}r^2 - m\ln r$.  Adding these results and using Eq.\ 
(\ref{solution}) gives the wavefunction $\psi_m$ of Eq.\ (\ref{basis}).
The current density of the tracking solution can be calculated using
Eq.\ (\ref{tv}), which yields Eq.\ (\ref{j_m}).  In the analysis it is seen that
the ${\bf A}^{\rm I}$ piece is responsible for the the radial part of the
polynomial prefactor in $\psi_m$, while the ${\bf A}^{\rm II}$ piece gives the
angular part.

In order to satisfy the tracking equation Eq.\ (\ref{track2D}) with energy
$E = \mu_B B_a$, we must introduce the scalar potential
\begin{equation}\label{Vsing}
   V({\bf r},m) = \mu_B b({\bf r}, m) = m \mu_B\phi_0 \delta({\bf r}).
\end{equation}
Since the wavefunction $\psi_m$ has a zero at the origin for $m > 0$, such a 
singular potential is absolutely harmless and it does not contribute to the 
physical properties of the system.  The appearance of the delta function is due 
to the transformation (\ref{Psi}), which transforms algebraic zeros of $\Psi$ 
into logarithmic singularities of $Q$, and the subsequent division  
of Eq.\ (\ref{schrodQ}) by the wavefunction to obtain the time
independent Riccati equation Eq.\ (\ref{eigS}).  Since $\nabla^2\psi_m$ is 
regular, the tracking solutions found also satisfy the original 
Schr\"odinger equation without the above $\delta$-function potential. 

It is of interest to point out that another tracking solution, associated
with the decomposition
\begin{equation}\label{AIAIIfp2}
   {\bf A}^{\rm I} = {\bf A}_a - {\bf a}({\bf r}, m),\qquad
   {\bf A}^{\rm II} =  -{\bf a}({\bf r}, m),
\end{equation}
yields the same LLL energy eigenvalue $E_m= \mu_B B_a$ and current density Eq.\ 
(\ref{j_m}), but with basis function $\bar{\psi}_m$ whose prefactor is {\em 
analytic}.  The form $\bar{\psi}_m$, which is the one used by Laughlin, 
\cite{laughlin2} corresponds to a vector potential ${\bf A} = {\bf A}_a - 2{\bf 
a}({\bf r}, m)$.  Hence there is a (negative) net flux tube attached to the 
origin of the coordinate system, which the moving electron observes and picks up 
an Aharonov-Bohm phase.

Now consider the $N$-electron problem. The construction (\ref{AIAIIfp}) can be 
generalized, using the particle-dependent decomposition
\begin{equation}
   {\bf A}^{\rm I}_i = {\bf A}_a - {\bf a}({\bf r}_i, m_i), \qquad
   {\bf A}^{\rm II}_i =  {\bf a}({\bf r}_i, m_i),
\end{equation}
with ${\bf a}$ defined as in Eq.\ (\ref{arm}). We find that
$Q =1/4 l_a^{-2}\sum_i Z_i\bar{Z}_i - \sum_i m_i\ln \bar{Z}_i$,
which leads to the eigenfunction
\begin{equation}
\Psi_{\{m\}}\! =\! \bar{Z}_1^{m_1}\bar{Z}_2^{m_2} \cdots \bar{Z}_N^{m_N}
           \exp\!\left(\!-{1\over 4}l_a^{-2}\sum_i Z_i\bar{Z}_i \right)\!.
\end{equation}
In the case of a fully filled LLL, $\nu=1$, for a circular droplet with unit 
area, the particle number is $N=B_a/\phi_0$, and the one-electron states are 
filled 
with increasing $m$ quantum numbers up to $N-1$. The total canonical angular 
momentum $M = \sum_{i=1}^N m_i = 0+1+\dots+(N-1)=N(N-1)/2$ is a conserved 
quantity.  
The angular momentum, however, can be arbitrarily distributed among the 
particles; 
thus any linear superposition
\begin{eqnarray}
\Psi_M &=& \sum_{\{m\}} c_{\{m\}} \Psi_{\{m\}},
\end{eqnarray}
subject to the constraint $\sum_i m_i=M$, is a valid solution. The
$N$-electron wavefunction must be antisymmetric with respect to the
exchange of the particles. This requires that $c_{\{m\}}$ be the totally
antisymmetric tensor of $N$ indices, i.e., we have to form a Slater
determinant. A closed form is achieved by forming the Vandermonde
determinant\cite{gradshteyn}
\begin{eqnarray}
   \sum_{\{m\}} c_{\{m\}} \prod_i \bar{Z}_i^{m_i} &=&
   \det\left|\begin{array}{lllll} 1 & \bar{Z}_1 & \bar{Z}_1^2 & \cdots & 
   \bar{Z}_1^{N-1}\\
                               1 & \bar{Z}_2 & \bar{Z}_2^2 & \cdots & 
   \bar{Z}_2^{N-1}\\
                               \vdots&\vdots&\vdots& &\vdots\\
                               1 & \bar{Z}_N & \bar{Z}_N^2 & \cdots & 
   \bar{Z}_N^{N-1}
          \end{array} \right| \nonumber\\
   &=&\prod_{i=1}^N\;\prod_{j>i}^N\; (\bar{Z}_j-\bar{Z}_i).
   \label{vand}
\end{eqnarray}
Thus the $N$-electron wavefunction for the fully filled LLL takes the form
\begin{eqnarray}\label{Psi_M}
   \Psi_M = \prod_{i=1}^N\prod_{j>i}^N\; (\bar{Z}_j-\bar{Z}_i)
   \exp\left(-{1\over 4}l_a^{-2}\sum_i Z_i\bar{Z}_i \right),
\end{eqnarray}
where $2M = N(N-1)$.

%--------------------------------------------------------------------------
\subsection{Interacting Particles - The Laughlin Wave Function}

The above discussion remains equally valid if the system is subject to
internal fields created by the particles themselves.  Our general technique to 
find solutions to the many body problem is to solve the one-body 
problem, form product wave functions from these solutions, and finally 
antisymmetrize them for fermions (or symmetrize for bosons).  

Consider the addition of a Chern-Simons vector potential, whose value at the 
position of the $i$th electron is determined by the positions of the other 
particles as
\begin{equation}
   {\bf a}_{cs}({\bf r}_i, \tilde{\phi} ) =  \tilde{\phi}{\phi_0\over 2\pi}
   \sum_{j\neq i}^N \hat{\bf z}\times {({\bf r}_i-{\bf r}_j)
                                \over |{\bf r}_i-{\bf r}_j|^2},
\end{equation}
with $\tilde{\phi} = \phi_{cs}/(\phi_0 N)$ the number of flux quanta per 
particle 
arising from ${\bf a}_{cs}$.  The Chern-Simons vector potential ${\bf a}_{cs}$ 
attaches fictitious flux tubes of density
\begin{equation}
   {\bf b}_{cs}({\bf r}_i, \tilde{\phi}) =  \nabla_i\times {\bf a}_{cs}({\bf 
   r}_i) 
   = \hat{z}\tilde{\phi}\phi_0 \sum_{j\ne i}^N \delta({\bf r}_i-{\bf r}_j) 
\end{equation}
to the particles. These composite particles remain fermions if
$\tilde{\phi}$ is even, while their statistics transmute to bosons when
$\tilde{\phi}$ is odd. 

As above, we decompose the vector potential ${\bf A}$ as
\begin{eqnarray}
   &&{\bf A}^{\rm I}_i = {\bf A}_a - {\bf a}({\bf r}_i, m) 
   - {\bf a}_{cs}({\bf r}_i,\tilde{\phi}), \nonumber\\
   &&{\bf A}^{\rm II}_i = {\bf a}({\bf r}_i, m) + {\bf a}_{cs}({\bf 
   r}_i,\tilde{\phi}).\label{AI-IIcs}
\end{eqnarray}
Again the flux tubes appear with opposite sign in ${\bf A}^{\rm I}$ and
${\bf A}^{\rm II}$, so they cancel out from the Hamiltonian. The associated 
eigenfunction is formed by solving Eq.\ (\ref{tr2D}) using Eq.\ (\ref{AI-IIcs}).
Equation (\ref{track2D}) is satisfied with energy $E=N\mu_B B_a$, provided 
that we introduce the fictitious scalar potential
\begin{equation}
   V({\bf r}_i, m ,\tilde{\phi}) = \mu_B\phi_0 [m\delta({\bf r}_i) +
   \tilde{\phi}\sum_{j\ne i}^N  \delta({\bf r}_i-{\bf r}_j)].
\end{equation}
Recalling the discussion after Eq.\ (\ref{Vsing}), this singular potential
is spurious and has no real effect on the problem.
Again, the wavefunctions to be found below are valid solutions of
the original Schr\"odinger equation without $V({\bf r}_i, m ,\tilde{\phi})$.

The Chern-Simmons contribution gives
\begin{eqnarray}
   \Im Q_{cs} &&= -\tilde{\phi}\sum_i\sum_{j\neq i}\tan^{-1}\left({y_i-y_j\over 
   x_i-x_j}\right), \nonumber\\
   \Re Q_{cs} &&= -\tilde{\phi}\sum_i\sum_{j\neq i}
                  \ln |{\bf r}_i - {\bf r}_j|,\nonumber
\end{eqnarray}
which, combined with the other contributions calculated in the free particle 
case, 
gives
\[
   Q\! = 1/4 l_a^{-2}\sum_i Z_i\bar{Z}_i -\sum_i  m_i \ln \bar{Z}_i - 
   \tilde{\phi} \sum_i\sum_{j>i} \ln (\bar{Z}_i-\bar{Z}_j).
\]
Following the anti-symmetrization procedure in Eq.\ (\ref{vand}), the 
wavefunction takes the form
\begin{eqnarray}
   \Psi_L &=& \prod_i\;\prod_{j>i}\;(\bar{Z}_j-\bar{Z}_i)^{\tilde{\phi}}
   \Psi_M\nonumber\\
   &=& \prod_i\;\prod_{j>i}\; (\bar{Z}_j-\bar{Z}_i)^{\tilde{\phi}+1}
   \exp\left(- {1\over 4}l_a^{-2}\sum_i Z_i\bar{Z}_i \right),\nonumber\\
   \label{Laugh}
\end{eqnarray}
where $\Psi_M$ is the free electron wave function of Eq.\ (\ref{Psi_M}).
This is the Laughlin wavefunction, used 
to describe the state of the electrons for fractional filling factors 
$1/(\tilde{\phi} + 1)$ (with $\tilde{\phi}$ even). The distinction
between type-I and type-II 
potentials is that the type-I piece generates the radial part of the Laughlin 
prefactor, while the type-II piece generates the angular part. It is the type-I 
vector potential in the above construction that keeps the particles further away 
from each other, and increases the effective filling factor to $\nu_{\rm 
eff}=1$.  

Our derivation of the Laughlin wavefunction deserves some interpretation.
It is well-known that without the repulsive interaction of the particles any
wavefunction with an anti-analytic prefactor is an eigenstate in the
LLL.\cite{Girvin}
The state in Eq.\ (\ref{Laugh}) is only one of the many degenerate ground 
states.  Switching on the weak Coulomb interaction slightly mixes up the 
different Landau levels. The Laughlin function $\Psi_L$ ceases to be an exact 
ground state but in the high field limit it is believed to become a good 
variational ground state, better, in general, than other wavefunctions with 
anti-analytic prefactors.

Usually, the Laughlin wavefunction is derived in a field theoretic framework 
beyond the mean field level. Starting either from a fermionic\cite{lopez} or 
bosonic\cite{zhang} field theory, fictitious Cherns-Simons flux tubes are
attached to the particles. Comparing these theories, based on singular gauge
transformations, to our decomposition of the vector potential into type-I
and type-II components, it is obvious that the former theories only take into
account the singular type-II piece. As a consequence, these mean-field theories 
are not able alone to account for the radial component of the Laughlin 
prefactor. The full Laughlin prefactor only emerges when fluctuations of the 
fictitious gauge 
field beyond the mean field level are taken into account. In contrast, our 
construction correctly accounts for fluctuations and produces the full Laughlin 
wave function. In this sense it goes beyond the mean field theory.

Another recent theory due to Rajaraman and Sondhi\cite{rajaraman} rather 
artificially attaches complex flux tubes to the electrons, using a {\em complex} 
Chern-Simons-type vector potential ${\bf a}_{cs}-i\hat{\bf z}\times{\bf 
a}_{cs}$. The imaginary term in the vector potential allows them to obtain the 
complete $\Psi_L$ at the mean field level. The Aharonov-Bohm phase picked up by 
an electron moving in the presence of such a complex vector potential 
is $Q_{\rm AB} = -i(2\pi/\phi_0)\int ({\bf a}_{cs} - i\hat{\bf z}\times{\bf 
a}_{cs})d{\bf r}$.  In our construction a moving particle observes {\em real} 
type-I and type-II Chern-Simons vector potentials, ${\bf A}^{\rm II}_{cs} =
-{\bf A}^{\rm I}_{cs} = {\bf a}_{cs}$. The type-I component gives the 
modification  $\Im Q = (2\pi/\phi_0)\int {\bf a}_{cs}d{\bf r}$ while the
type-II component gives $\Re Q= (2\pi/\phi_0)\int (\hat{\bf z}\times{\bf 
a}_{cs})d{\bf r}$.  Since $\Re Q + i\Im Q = -Q_{\rm AB}$, the two approaches 
give the same mathematical wave function.  However the physics is different: 
our approach does not attach bare flux tubes to the particles. The full 
wavefunction is constructed by assigning a different role to the two 
components of the vector potential.  We use the Chern-Simons vector 
potential ${\bf a}_{cs}$ only as an
auxiliary tool in deriving new solutions of the Schr\"odinger equation.
The term ${\bf a}_{cs}$ cancels out from the Hamiltonian, in which the total 
vector potential is ${\bf A}_a$, due only to the uniform applied field. On the 
other hand, the complex vector potential used in Ref.\ \onlinecite{rajaraman} 
is an unrealistic characteristic which remains in the Hamiltonian.

\subsection{Coulomb Tracking}

In previous subsections we analyzed problems with a uniform applied 
magnetic flux density $B_a$. Now we consider an additional one-body, 
non-uniform flux density $B_1({\bf r})$, giving the total flux density 
\begin{equation}
   B({\bf r}) = B_a - B_1({\bf r}).
\end{equation}
Choosing the associated vector potential in the Coulomb gauge, $\nabla\cdot{\bf 
A} = 0$, allows us to interpret $B({\bf r})$ as
generated by a type-I vector potential. In general, non-uniform perturbations
of this kind lift the degeneracy of the LLL, and the one-electron eigenfunctions
cannot be found analytically.  However, according to Eq.\ (\ref{track2D}), if 
$B({\bf r})$ is tracked by a scalar potential $V_1({\bf r})$, i.e.
\begin{equation}
   \mu_B B_1({\bf r}) = V_1({\bf r}) ,
   \label{pert}
\end{equation}
the degeneracy of the LLL is preserved.  Since any nonzero spatial average 
components of the flux density and scalar potential only have a trivial level 
shifting effect on the LL structure, we assume that the mean values of 
$B_1$ and $V_1$ are zero in the following treatment.  
Although the LLL energy does not 
change, the basis functions spanning this level pick up an extra prefactor
\begin{eqnarray}
   \psi_m^\prime({\bf r})&=& e^{-\Re Q_1}\psi_m({\bf r}),
   \label{newbasis}
\end{eqnarray}
where $\Re Q_1({\bf r})$ is the solution of the real part of Eq.\ 
(\ref{tr2D}) with ${\bf A}^{\rm I}$ the solution of 
$\nabla\times{\bf A}^{\rm I} = -B_1({\bf r})$.

This result for the single electron states was pointed out in 
Ref.\ \onlinecite{dubrovin}, although 
all explicit analysis was based on a Pauli 
spinor Hamiltonian with no scalar potential.  However, we proceed further and 
suppose that the perturbing magnetic field is due to a two-body, type-I vector 
potential in a system of $N$ electrons. Our goal is to construct a Hamiltonian, 
which involves realistic two-body scalar potentials, such as the screened 
Coulomb interaction,
\begin{equation}\label{scCoul}
   V_1({\bf r}_i) = g_c e^2\sum_{j\ne i}
   \frac{e^{-|{\bf r}_i - {\bf r}_j|/\xi}} {|{\bf r}_i - {\bf r}_j|},
   \qquad 0 \le g_c \le 1,
\end{equation}
whose ground state can be constructed exactly. The interaction is characterized 
by an exponential screening parameter $\xi$ and a linear strength parameter 
$g_c$, with nominal value $g_c = 1$.  According to the tracking
equation Eq.\ (\ref{track2D}) this requires the presence of a screened magnetic
monopole-like field 
\begin{equation}
    B_1({\bf r}_i) = {g_m e^2\over\mu_B}\sum_{j\ne i}
                    \frac{e^{-|{\bf r}_i - {\bf r}_j|/\xi}}
                         {|{\bf r}_i - {\bf r}_j|},
   \label{magn2b}
\end{equation}
with strength $g_m = g_c$. With this proviso the degeneracy of the LLL remains 
intact and the ground state wavefunctions have the form
\begin{eqnarray}
   \Psi^\prime({\bf r})&=& e^{-\Re Q_1}\Psi({\bf r}),
   \label{newsol}
\end{eqnarray}
where $\Re Q_1(\{{\bf r}\})$ is again the solution of the real part of
Eq.\ (\ref{tr2D}) with $\nabla_i\times{\bf A}^{\rm I} = -B_1({\bf r}_i)$,
and $\Psi$ is the wave function in the absence of $V_1$. In the present case,
with $B_1$ given by (\ref{magn2b}), one obtains
\begin{equation}\label{QCscreen}
   \Re Q_1(\{{\bf r}\})\! =\! \alpha\xi\sum_i\sum_{j>i}
   \left[{\rm Ei}\!\left(\!-\frac{|{\bf r}_i - {\bf r}_j|}{\xi}\right)\! -\!
   \ln\!\left(\frac{|{\bf r}_i - {\bf r}_j|}{\xi}\right)\right]\!\!,
\end{equation}
where ${\rm Ei}(r)$ is the exponential integral function ${\rm Ei}(r)=
-\int_{-r}^\infty dt\exp(-t)/t$, and $\alpha = 2m^*e^2g_c/\hbar^2$.  In the 
strict Coulomb limit $\xi\to\infty$, Eq.\ (\ref{QCscreen}) reduces to 
\begin{eqnarray}
   \Re Q_1(\{{\bf r}\}) = -\alpha\sum_i\sum_{j>i}|{\bf r}_i - {\bf r}_j| .
   \label{strict}
\end{eqnarray}
Assuming tacitly that the ground state structure does not change
drastically in the presence of a finite electron-electron interaction, i.e.,
the system remains in the Laughlin liquid phase and does not cross
over into a Wigner solid,\cite{price} 
we pick up the Laughlin wavefunction from the set of degenerate
wavefunctions of the LLL, and set $\Psi=\Psi_L$ in Eq.\ (\ref{newsol}).
In the pure Coulomb case, using Eq.\ (\ref{strict}), this leads to the
Coulomb tracking wave function
\begin{eqnarray}
   \Psi_C = &&\prod_i^N\;\prod_{j>i}^N\; (\bar{Z}_j-\bar{Z}_i)^{\tilde{\phi}+1}\,
            \exp\left(\alpha\sum_i\sum_{j>i}|{\bf r}_i - {\bf r}_j|\right.
            \nonumber\\
            &&\left. -{1\over 4}l_a^{-2}\sum_i |{\bf r}_i|^2\right).
\end{eqnarray}

The magnetic flux density in Eq.\ (\ref{magn2b}) may appear rather artificial, 
but it can be viewed as a smeared out analog of the standard Chern-Simons flux 
tube.  In comparison with the Laughlin wavefunction the extra Jastrow factor
$\exp(-\Re Q_1)$ pushes the particles further apart, decreasing even more the
probability of two particles approaching each other, and thus decreasing the 
Coulomb energy. However, this factor does not modify the effective filling 
factor: since $\Re Q_1$ appears in the exponent, it does not change the 
polynomial prefactor.  In a realistic situation with a Coulomb interaction 
between
the particles, $\Psi_C$ may be a useful variational ground state wavefunction.
If one uses Eq.\ (\ref{QCscreen}), the corresponding screened Coulomb 
tracking function depends on two parameters $\xi$ and $g_c$, which
can be adjusted variationally, independently of their nominal values in Eq.\ 
(\ref{scCoul}). 

\begin{figure}[hbt]
\epsfxsize=\columnwidth\epsfbox{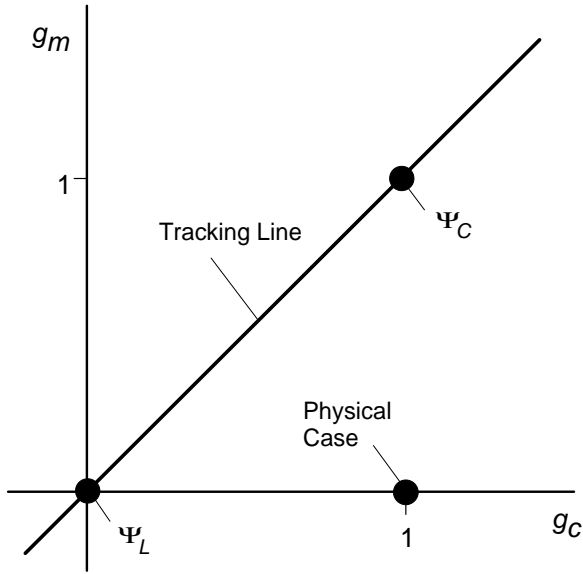} 
\caption{For the model parameter space shown, $g_c$ and $g_m$ are the strengths 
of the Coulomb and tracking magnetic field interactions, respectively.  Exact 
tracking solutions in the space of two-body couplings exist along the tracking 
line, $g_m = g_c$. Laughlin's wavefunction $\Psi_L$ is an exact solution
at $g_m = g_c = 0$. The real physical problem is characterized by $g_c = 1$,
and $g_m = 0$. Any point on the tracking line could be used as a variational
trial state.} 
\label{fig} 
\end{figure}

Figure \ref{fig} illustrates schematically the situation in the 
space of model parameters: Exact ground state wavefunctions can be constructed
along the straight line denoting models with equal strengths of the
Coulomb interaction and its tracking two-body magnetic field. Laughlin's
wavefunction $\Psi_L$ refers to the special point $g_m = g_c = 0$, while the
full Coulomb tracking wavefunction $\Psi_C$ refers to the point $g_m =g_c =1$. 
A priori, $\Psi_L$ and $\Psi_C$ are equally good (or bad) 
representations of the 
wave function for the the real physical case $g_c = 1$, $g_m = 0$. Improvement 
on the FQHE ground state could be achieved by considering tracking solutions 
where the
common value of the coupling constants $g_m = g_c = g$ is used as a variational
parameter, which refers to moving along the tracking line of the figure.

The prefactor of $\Psi_C$, including the real factor $\exp(-\Re Q_1)$, is not 
completely anti-analytic; thus it is not entirely in the lowest Landau level.  
Nevertheless, it has presumably an extensive overlap with the 
LLL due to the dominating anti-analytic polynomial factor. In a situation where 
the Coulomb energy is comparable to the magnetic energy, Landau level mixing
is not negligible, and $\Psi_C$ could be tested directly as a variational
ground state. Note that the form of the Jastrow correction factor $\exp(-\Re 
Q_1)$ taken from Eq.\ (\ref{QCscreen}) or (\ref{strict}) is different from
the form used recently by Price {\em et al}.\cite{price} 
It would be interesting to
calculate numerically the variational upper bound on the ground state energy
set by the tracking solution and compare it directly to that of
Ref.\ \onlinecite{price}.

%==============================================================================
\section{Conclusions}

In summary, we analyzed a system of $N$ electrons interacting with
externally applied non-uniform scalar and vector potentials, and with
each other through two-body potentials. Our approach was based on
transforming the Schr\"odinger equation into a nonlinear Riccati
equation and then linearizing the latter with an appropriate ansatz.
We showed that when the magnetic flux density tracks the spatial dependence
of a scalar potential, exact analytic solutions of the Schr\"{o}dinger equation
can be obtained in any dimension.  These ``tracking'' solutions, which form a 
Jastrow product, are characterized by a unit vector $\hat{\bf n}$, and a 
decomposition of the vector potential.  In the 2D examples analyzed here, the 
tracking solutions constitute the degenerate ground states of the system. In 
particular, the tracking solution corresponding to the 2D electron gas in a
homogeneous transverse magnetic field was found to be the Laughlin wavefunction
whenever the vector potential decomposition is based on the Chern-Simons vector 
potential ${\bf a}_{cs}$. Our construction, however, does not attach
singular flux-tubes to the particles, since ${\bf a}_{cs}$ does not appear in 
the Hamiltonian.

Adding a realistic repulsive two-body potential, such
as the screened Coulomb potential, a 
tracking solution exists only if there is a 
two-body magnetic field present.  Nevertheless, the resulting tracking wave 
function may be useful in a variational calculation to provide further insight 
into the nature of the fractional quantum Hall ground state in the presence of 
strong electron-electron interactions and Landau level mixing.  

In three dimensions there are many possible tracking solutions not considered 
here. Furthermore, the tracking constructions that linearize the Riccati 
eigenequation also linearize the time-dependent Riccati equation, creating 
another domain of exactly solvable problems. 
We leave these interesting problems to future study.

\acknowledgments
Discussions with P. Erd\H{o}s and J-J.\ Loeffel are appreciated, and 
the partial support of the Swiss National Science Foundation through 
grant No. 20-46676.96 is acknowledged.
 
%========================================================================

\end{document}